\documentclass[aps,prb,twocolumn,groupedaddress,a4paper,byrevtex,showpacs]{revtex4}

\usepackage{graphicx}
\usepackage{amsmath}
\usepackage{amssymb}

\begin{document}


\title{Evidence of a dominant Maki-Thompson process in the
fluctuation conductivity of Bi-2212 superconducting whiskers}



\author{M.~Truccato}
\email[Electronic address: ]{truccato@to.infn.it}
\affiliation{INFM UdR Torino Universit\`a - Dip. Fisica Sperimentale - Via P. Giuria 1, I-10125, Torino, Italy}

\author{G.~Rinaudo}
\affiliation{INFM UdR Torino Universit\`a - Dip. Fisica Sperimentale - Via P. Giuria 1, I-10125, Torino, Italy}

\author{A.~Causio}
\affiliation{INFM UdR Torino Universit\`a - Dip. Fisica Sperimentale - Via P. Giuria 1, I-10125, Torino, Italy}

\author{C.~Paolini}
\affiliation{INFM UdR Torino Universit\`a - Dip. Fisica Sperimentale - Via P. Giuria 1, I-10125, Torino, Italy}

\author{P.~Olivero}
\affiliation{INFM UdR Torino Universit\`a - Dip. Fisica Sperimentale - Via P. Giuria 1, I-10125, Torino, Italy}

\author{A.~Agostino}
\affiliation{INFM UdR Torino Universit\`a - Dip. Chimica Generale ed Organica Applicata - C.$^{so}$ Massimo D'Azeglio 48, I-10125, Torino, Italy}



\date{Submitted to Physical Review B on September 27$^{th}$, 2004, revised on March 21$^{th}$, 2005}

\begin{abstract}
We report the measurement of the \textit{a}-axis fluctuation conductivity in zero field
for Bi$_{2}$Sr$_{2}$CaCu$_{2}$O$_{8+x}$ microcrystals. A good geometrical characterization and an original method
for data analysis allow to determine its temperature behaviour to a high
accuracy. 
The correct application of the  
complete fluctuation theory
[A. A. Varlamov \textit{et al.}, Adv. Phys. \textbf{48}, 655 (1999)], with
a minor correction for the $\tilde k$ factor, leads to the conclusion that the anomalous MT term is the
most important contribution throughout the whole temperature range of interest. 
More traditional alternative interpretations of the experimental data are also investigated and their 
inconsistency is demonstrated.
This implies an
\textit{s}-wave symmetry for the order parameter and could provide a physical mechanism
for the preformed pair scenario.
\end{abstract}



\maketitle



\section{\label{Intro}INTRODUCTION}
The additional contribution to the normal state conductivity due to Cooper pairs
formed above $T_{c}$ ($\Delta\sigma$, also called paraconductivity)
was theoretically investigated for the first time by Aslamazov and Larkin (AL),\cite{AL_1968}
who obtained for 2D systems
$\Delta \sigma_{\text{AL}}^{\text{2D}} = \frac{e^{2}}{16\hbar d}\varepsilon ^{-1}$,
where $\varepsilon=\frac{T-T_{c}}{T_{c}}$ and $d$ is the thickness of
the 2D layer. Soon after, Maki and Thompson (MT)\cite{Maki_1968,Thompson_1970}
calculated other two leading contributions (one regular and one anomalous) to the paraconductivity,
showing that decayed pairs can scatter on an impurity potential as much as they were still paired.
These contributions explain fairly well experimental data for low $T_{c}$ superconductors,\cite{Skocpol_Tinkham_1975}
but the problem is still open for high $T_{c}$ materials.
From the theoretical point of view, Dorin \textit{et al.} \cite{Dorin_1993} pointed out the necessity of
taking into account also the fact that electrons involved in fluctuating pairs cannot take part in ordinary
conductivity, therefore suppressing the quasiparticle density of states (DOS).
Recently, a comprehensive model consisting
of the AL, the MT and the DOS terms was formulated, which includes also a slight modification in
the numerical coefficient for the MT anomalous term. \cite{Varlamov_1999,Larkin_2004}
On the experimental side, the relative importance of each contribution, especially of the MT terms,
has long been debated. The leading role of the AL contribution is presently generally accepted and the DOS term
is believed to be a comparable correction. About the MT terms, their contribution was
reliably investigated only recently, after that the importance of the DOS term was completely understood.\cite{Axnas_1998}
These experiments, performed both in a magnetic field \cite{Wahl_1999,Wahl_1999_2,Thopart_2000,Bjornangen_2001,Kim_2003}
and in zero field,\cite{Chowdhury_1999,Kim_2003} share the conclusion that the MT process is unimportant.
The purpose of the present work is to show that an accurate analysis of the
zero-field paraconductivity measured in ideal experimental conditions
can lead to very different conclusions, setting the anomalous MT term
as the most important throughout the whole temperature range of validity
of the superconducting fluctuation model.
It is worth noting
that the anomalous MT term is the only one whose presence or absence could have implications about the
symmetry of the order parameter wavefunction, due to the fact that it
cannot exist in the case of a pure \textit{d}-wave symmetry.\cite{Carretta_1996}

\section{\label{Exp}EXPERIMENTAL}
Our experiment was performed by using microscopic whisker-like single crystals of
Bi$_{2}$Sr$_{2}$CaCu$_{2}$O$_{8+x}$ (Bi-2212) grown by the method of the glassy
plates oxygenation.
These crystals grow with the
\textit{a}-axis aligned with the length direction, the \textit{b}-axis in the
width direction and the \textit{c}-axis along the thickness. Typical crystal sizes are
1 mm in length, 10 $\mu$m in width and 1 $\mu$m in thickness and it has already been demonstrated that
this kind of samples shows a considerably lower density of defects in comparison with bulk ones.\cite{Gorlova_95,Timofeev_98}
The fact that in Bi-2212 the oxygen diffusion occurs essentially in the \textit{ab}-plane with the highest
rate along the \textit{a}-axis direction,\cite{Li_1994} which is also the direction with the largest crystal size,
makes it very likely that the small volumes intended for the investigation of the excess conductivity
in our experiment (maximum size of about $300 \times 6 \times 1 \ \mu$m$^{3}$) are homogeneous from the
point of view of the oxygen doping or, at least, more homogeneous than the volumes tested
in experiments on bulk samples.

Eleven crystals were mounted onto sapphire substrates
and silver plus gold electrodes were evaporated on their top surfaces to obtain a chip suitable
for standard four-probe resistivity measurements.
The details of the sample preparation have been reported in a previous paper \cite{Truccato_02} and resulted
in a typical contact resistance $\approx 2 \Omega$, corresponding to a specific resistance
$\approx 10^{-6}$ $\Omega \cdot \text{cm}^2$.

Since we were interested in the absolute measurement of the paraconductivity, a lot of attention was paid
to the sample geometrical characterization. All sizes in the length direction were measured by means of SEM
comparisons with calibrated standards, while crystal widths and thicknesses were measured by scanning the crystal
length in a series of AFM maps, each of them about $22\times 22 \mu\text{m}^{2}$ large.
This procedure, along with the
electrical characterization, allowed to select very regularly-shaped single phase crystals
for further analysis. At the end of this stage, only 2 nearly perfect samples survived.
Their sizes are reported in Table~\ref{Tab_1}: one sample (WI3C\_1) shows a 78 nm growth step,
so that its cross section corresponds to two neighbouring rectangles.
It should be noted that the uncertainties
in the thickness are quite large, even if care
was taken to exclude the regions corresponding to the submicron droplet-like impurities
located on the top of the crystals.\cite{Truccato_02}
Actually, the uncertainties in
the sample geometry $(\approx 10 \%)$
are the major sources of error for this study and a special procedure was
developed to overcome this problem, as it will be explained below.

%

\begin{table} [htbp]  
\caption{\label{Tab_1}Relevant sizes for samples WI3B\_1 and WI3C\_1. $L$ is the total crystal length,
$\Delta x$ the distance between midpoints of voltage contacts, $W$ is the crystal width and $t$ the thickness.
WI3C\_1(\textit{a}) and  WI3C\_1 (\textit{b}) refer to the two parts in which the growth step divides longitudinally
the crystal (see text).}
\begin{ruledtabular}
\begin{tabular}{ccccc}
Sample                & $L(\mu \text{m})$   & $\Delta x (\mu \text{m})$ & $W(\mu \text{m})$ & $t(\mu \text{m})$  \\  \hline
WI3B\_1               &  1486.3$\pm$2.8  &   300$\pm$3          &  6.35 $\pm$0.26  &   1.10$\pm$0.08   \\
WI3C\_1(\textit{a})   &   1499$\pm$23    &   129$\pm$8         &    3.26$\pm$0.15  & 0.680$\pm$0.021    \\
WI3C\_1 (\textit{b})  &                  &                     &   2.86$\pm$0.13   & 0.758 $\pm$ 0.024 \\
\end{tabular}
\end{ruledtabular}
\end{table}

\begin{figure}[htbp]
\includegraphics[angle=270,width=8.2 cm]{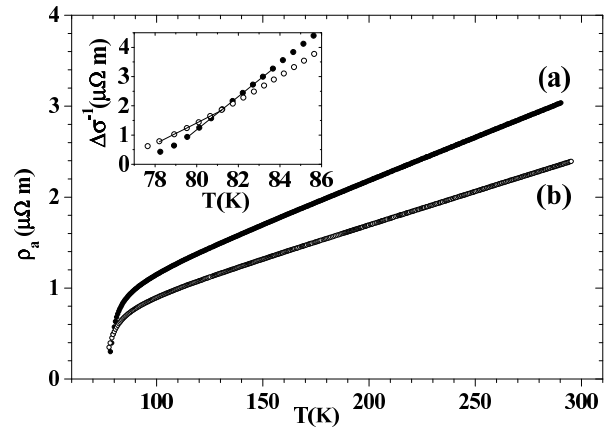}
\caption{ \label{fig.1_PRB} Temperature behaviour of the resistivity $\rho_{a}$
in the \textit{a}-axis direction for sample WI3B\_1 (curve (a), solid circles)
and WI3C\_1 (curve (b), open circles). The feeding currents were $I=1\mu$A for sample WI3B\_1 and $I=2\mu$A for
sample WI3C\_1. The vertical error bars associated to each data point are not shown for clarity.
The inset shows the reciprocal of the
paraconductivity $\Delta \sigma$ as a function of the temperature. Solid lines represent
the fits to the pure 2D AL formula.}
\end{figure}

Fig.~\ref{fig.1_PRB} shows the temperature behaviour of the
resistivity along the \textit{a}-axis $(\rho_{a})$ for the two samples.
The resistivity was obtained by four-probe voltage measurements considering
the distance between the electrode midpoints. In fact, the analytical treatment by Esposito
\textit{et al.} \cite{Esposito_2000}
has shown that the disentaglement of the in-plane ($\rho_{ab}$) and out-of-plane ($\rho_{c}$)
resistivity components occurs for
$L \gg \pi \sqrt{\rho_{c}/\rho_{ab}} \; t$,
which is within our experimental range, and the corresponding expression results
in relative corrections for our samples below $4 \cdot 10^{-7}$, that can be neglected.
The measurements were collected on sample warming in steps of 0.5 K every 130 s;
no care was put in screening the Earth's magnetic field. The average of the readings collected in 10
subsequent current reversals at the rate of $\approx$ 1 Hz was performed in order to cancel the thermal
\textit{emf}'s,
with a temperature stability for the sample of $\pm0.01$ K during each reading cycle.
A conservative estimate of the absolute voltage error gave 60 nV, corresponding to a maximum relative error of about 0.2\%
in the temperature range used for the paraconductivity analysis.

\section{\label{analysis}DATA ANALYSIS AND DISCUSSION}
The first step in studying the paraconductivity
$\Delta \sigma \left( T \right) = 1/\rho_{a}(T)- 1/\rho^{n}_{a}(T)$
is the determination of the normal resistivity $\rho^{n}_{a}(T)$, which
is usually supposed to be linear in $T$. The validity of this behaviour down to $T_{c}$ is suggested by the
value of the Debye temperature $\Theta_{D}\approx255$K for Bi-2212.\cite{Stupp_1992} This implies a ratio
$T_{c}/\Theta_{D}\approx0.3$, which corresponds to a temperature range where no deviation
from linearity is observed in the most common metals.
The problems involved in this assumption have been often overcome by the measurement of the in-field paraconductivity
$\Delta \sigma \left( B,T \right) = 1/\rho_{a}(B,T)- 1/\rho_{a}(0,T)$,
\cite{Bjornangen_2001,Kim_2003,Thopart_2000,Holm_1995,Axnas_1998,Wahl_1999_2,Wahl_1999,
Latyshev_1995,Heine_1999,Pomar_1996,Balestrino_2001,Semba_1997}
but in principle a drawback in this approach
is the possible presence of ordinary magnetoresistance effects, which add to superconducting fluctuations.
\cite{Semba_1997} On the other hand, it has been recently reported \cite{Kim_2003} that the simple
assumption of a linear $T$ dependence leads to the same conclusions.
In our resistivity data no deviation from the linearity was detected down to $T \approx 160$ K, below which only
a downwards curvature was observable. For prudential reasons, we
chose to define the normal resistivity $\rho^{n}_{a}(T)$
as the low temperature extrapolation of the best linear fit for the experimental data corresponding to $T > 210$
K. All the analyses were repeated also for linear fits down to 160 K and, consistently, no significant
difference was found.
The part of the paraconductivity data corresponding to the temperature range near $T_{c}$ is shown in the
inset of Fig.~\ref{fig.1_PRB} in the form of its reciprocal
$\Delta \sigma^{-1}(T)$.

The experimental data were fitted to the full fluctuation theory in zero field:\cite{Varlamov_1999,Larkin_2004}
\begin{equation}
\Delta \sigma = \Delta \sigma_{\text{AL}} + \Delta \sigma_{\text{DOS}}+
\Delta \sigma_{\text{MT}}^{ \text{(an)} } + \Delta \sigma_{ \text{MT} }^{ \text{(reg)} } \quad , \label{DS_total}
\end{equation}
 where
\begin{eqnarray}
\Delta \sigma_{ \text{AL} } & = & \frac{e^{2}}{16 \hbar s}
\frac{1}{  \sqrt{\varepsilon \left( \varepsilon  + r \right)}     }   \label{AL} \\
\Delta \sigma_{ \text{DOS} } & = &  - \frac{e^{2}} {2 \hbar s} k  \ln \left( \frac{2} { \sqrt{\varepsilon}  +
\sqrt {\varepsilon + r} } \right)  \label{DOS} \\
\Delta \sigma_{\text{MT}}^{ \text{(an)} } & = & \frac{e^{2}}{4 \hbar s \left( \varepsilon - \gamma _{\phi} \right)}
\ln \left( \frac{\sqrt {\varepsilon} + \sqrt {\varepsilon  + r}} { \sqrt{\gamma _{\phi}}  +
\sqrt{\gamma _{\phi} + r} } \right)   \label{MT_an} \\
\Delta \sigma_{ \text{MT} }^{ \text{(reg)} } & = & - \frac{e^{2}} {2 \hbar s} {\tilde k}
\ln \left( \frac{2} { \sqrt{\varepsilon}  + \sqrt {\varepsilon + r} } \right) \ . \label{MT_reg}
\end{eqnarray}

Here $s$ is the spacing of the 2D superconducting layers in the \textit{c}-axis direction,
$\varepsilon = \ln \left( T/T_{c}\right)$ is the reduced temperature, $r\left( T \right)=-
{2J^2 \tau ^2 a}/{\hbar ^2}$ is the anisotropy
parameter, $\gamma _\phi \left( T \right) = - \tau a / \tau _\phi$ is the phase-breaking rate and
\[
k\left( T \right) = \frac{{ - \psi '\left( {\frac{1}{2} +
\frac{\hbar }{{4\pi  \cdot \tau k_B T}}} \right) + \frac{\hbar
}{{2\pi  \cdot \tau k_B T}}\psi ''\left( {\frac{1}{2}}
\right)}}  {\pi^{2} a }  \  ,
\]
with $J$ as the hopping energy between neighbouring layers,
$\tau$ the quasiparticle scattering time, $\tau_\phi$ the phase-breaking time
and $a$ given by the equation:
\[
a\left( T \right) =  \left[
{\psi \left( {\frac{1}{2} + \frac{\hbar }{{4\pi  \tau k_B
T}}} \right) - \psi \left( {\frac{1}{2}} \right) - \frac{\hbar
}{{4\pi \tau k_B T}}\psi '\left( {\frac{1}{2}} \right)}
\right] .
\]
For $\tilde k$ we used the value calculated in Ref.~\onlinecite{Larkin_2004}:
\[
\tilde k\left( T \right) = \frac{{ - \psi '\left( {\frac{1}{2} +
\frac{\hbar }{{4\pi \tau k_B T}}} \right) + \psi '\left(
{\frac{1}{2}} \right) + \frac{\hbar }{{2\pi  \tau k_B
T}}\psi ''\left( {\frac{1}{2}} \right)}}{\pi^{2} a } \ .
\]

The only independent parameters of the theory are $s$, $J$, $\tau$ and $\tau_\phi$;
besides, also the effect of the variation of the $T_{c}$ value must be carefully examined,
to take into account the broadness of the transition.

In order to obtain a first estimate of $T_{c}$, instead of the mathematically complex model described above, we
used the pure 2D AL law to fit the $\Delta \sigma^{-1}(T)$ data near $T_{c}$.
Since $\left( \Delta \sigma_{\text{AL}}^{\text{2D}} \right)^{-1} = 0$ for $T=T_{c}$, we extrapolated to
zero the best fits shown in the inset of Fig.~\ref{fig.1_PRB}, obtaining $T_{c}=77.9 \pm 0.6$ K
and $T_{c}=76.1 \pm0.6$ K for samples WI3B\_1 and WI3C\_1, respectively. Incidentally, one can note that
in this simple model the slope of $\Delta \sigma^{-1}(T)$ is related to the thickness $d$ of the isolated
2D superconducting layer, so that $d$ can be deduced from the experimental data. This gives $d=6.7 \pm 0.6$
\AA~and $d=4.1 \pm 0.4$ \AA~for sample WI3B\_1 and WI3C\_1, respectively, implying a sample-dependent thickness
of the superconducting layer.

Coming back to the full fluctuation theory, it should be noted that 
the identification of a proper random error with the minimum possible size is crucial,
in order to achieve the necessary numerical sensitivity in the analysis.
Therefore the analysis was not performed on the paraconductivity itself
$\Delta \sigma \left( T \right)$, which has a large relative error of about 10 \%, almost
completely due to the geometrical uncertainties and therefore temperature independent,
but on the non-normalized paraconductivity
$\Delta \sigma' \left( T \right) = \Delta \sigma  \cdot \frac{S}{I \Delta x}$ ($S$ is the sample cross section),
which shows a random relative error varying between 0.5\% and 1.3\% in the temperature
range used for the analysis. Only subsequently the results were translated to the values for $\Delta \sigma$,
when necessary.
It should be noted that, strictly speaking, the temperature behaviour of $\Delta \sigma' \left( T \right)$
is not exactly the same as $\Delta \sigma \left( T \right)$ because of the temperature dependences of $S$
and $\Delta x$, which are due to the thermal expansion. However, an estimate
based on the linear expansion coefficients measured in Ref.~\onlinecite{Meingast_1994}
gives for the relative difference between the two quantites a maximum value of 0.15 \% throughout 
the temperature range used in the analysis. Therefore it can be stated that,
within our experimental accuracy, $\Delta \sigma' \left( T \right)$ retains all the information
concerning the temperature behaviour of $\Delta \sigma \left( T \right)$.

The preliminary trials showed both that
the dirty limit $\left( 4 \pi \tau k_{B} T / \hbar \ll 1 \right)$ was not self-consistent and that the clean
limit $\left( 4 \pi \tau k_{B} T / \hbar \gg 1 \right)$ was not fully developed, so that the general
intermediate case had to be implemented, which required the tabulation of the digamma function
$\psi(x)$ in steps of $10^{-4}$.

\begin{figure}[htbp]
\includegraphics[width=7 cm]{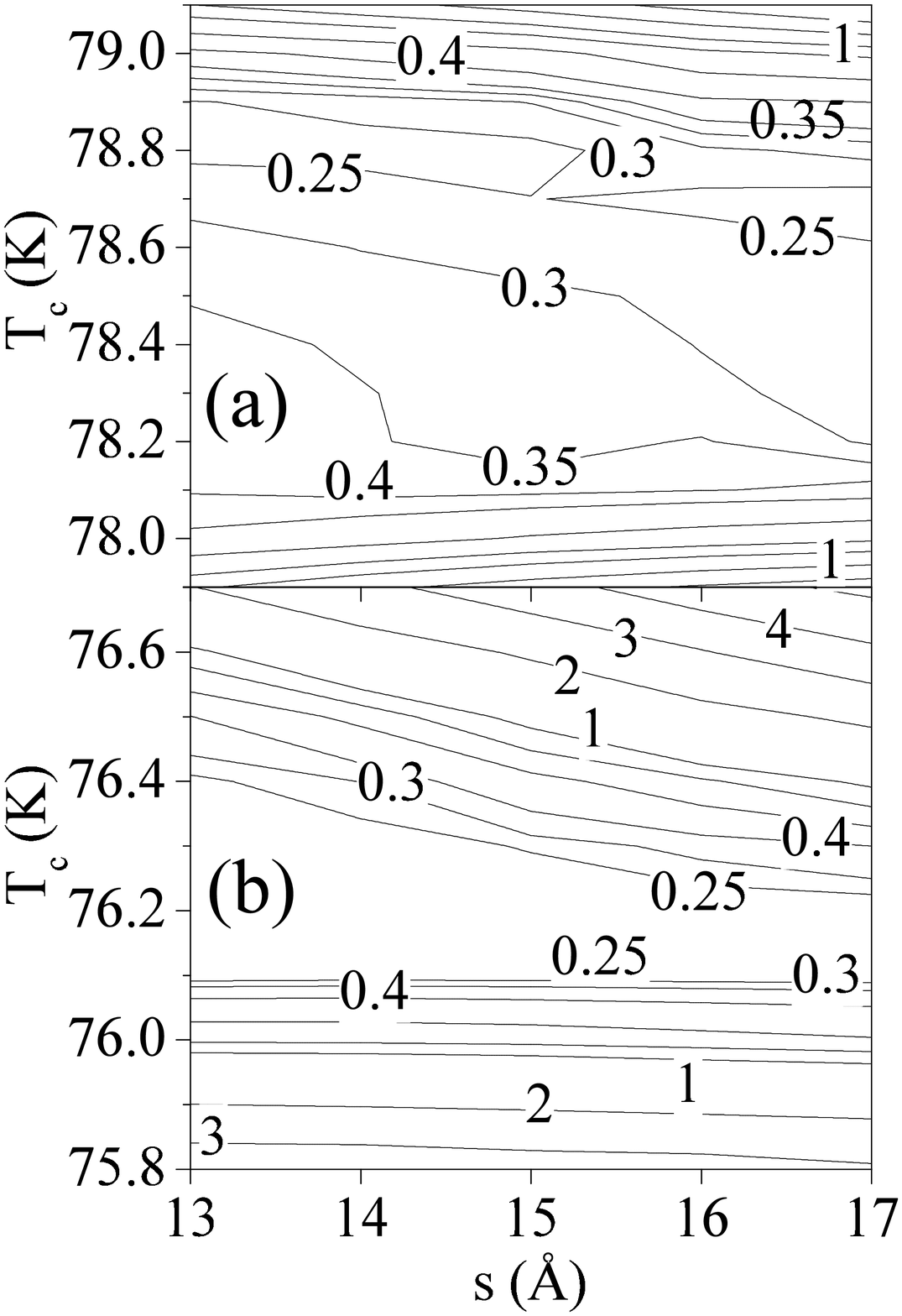}
\caption{ \label{fig_2} Contour plots of the reduced chisquare $\tilde \chi ^{2}=\chi ^{2}/\text{d.o.f.}$ in
the $(s, T_{c})$ plane for sample WI3B\_1 (a) and WI3C\_1 (b).
$\text{d.o.f.}=18$ for both samples.
Contiguous lines are variably spaced.
}
\end{figure}

By inspecting the paraconductivity expressions shown in Eqs.~(\ref{AL})--(\ref{MT_reg}), it can be noted that
$T_{c}$ acts as a normalization factor for the abscissa axis through the definition of $\varepsilon$, while $s$
does the same for the ordinate axis, since it can be factorized in all the terms. This suggests a
correlation between $s$ and $T_{c}$, implying that a good procedure for the fits consists in fixing
\textit{a priori} these two scale factors while leaving all the other parameters as free.
Therefore we selected a proper temperature range of the experimental data, fixed 
$s$ close to the expected value and $T_{c}$ close to the estimated value, and then performed the fit to 
$\Delta \sigma' \left( T \right)$ determining the best fit values of the $J$, $\tau$
and $\tau_\phi$ parameters, together with the corresponding minimum value of the reduced chisquare $\tilde \chi^{2}$.
By systematically varying $s$ and $T_{c}$,
this procedure was able to obtain
the contour lines of the minimum $\tilde \chi^{2}$ 
and of the best fit values of
$J$, $\tau$ and 
$\tau_\phi$ in the $(s,T_{c})$ plane for each sample.

The MINUIT routines were used to carry out the $\tilde \chi ^{2}$ minimization
by means of both the gradient and the Monte Carlo method, in order to be aware of both the local and the absolute
minima in the parameter space. Actually, when two practically equivalent $\tilde \chi^{2}$ minima were present,
only the one corresponding to the shortest $\tau_\phi$ was selected for conservative reasons, so that all the
fits satisfy the double criterium of both the least $\tilde \chi^{2}$ and the least $\tau_\phi$.

The contour plots for $\tilde \chi^{2}$ are shown in Fig.~\ref{fig_2}. It is apparent that both samples show
a wide minimum which is much more sensitive to the variations of $T_{c}$ than of $s$.

\begin{figure}[htbp]
\includegraphics[width=7 cm]{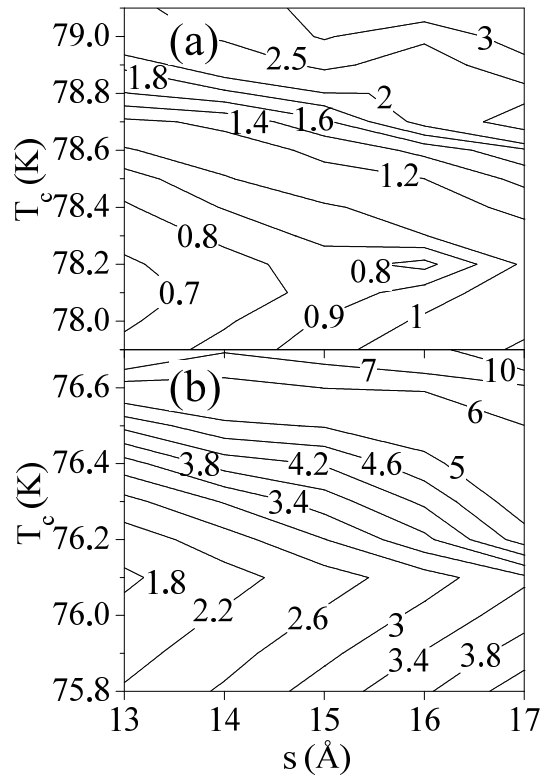}
\caption{ \label{fig_3} Contour plots of the phase-breaking time $\tau_{\phi}$ in
the $(s, T_{c})$ plane for sample WI3B\_1 (a) and WI3C\_1 (b). Labels are in units of $10^{-12}$ s.}
\end{figure}

\begin{figure}[htbp]
\includegraphics[width=7 cm]{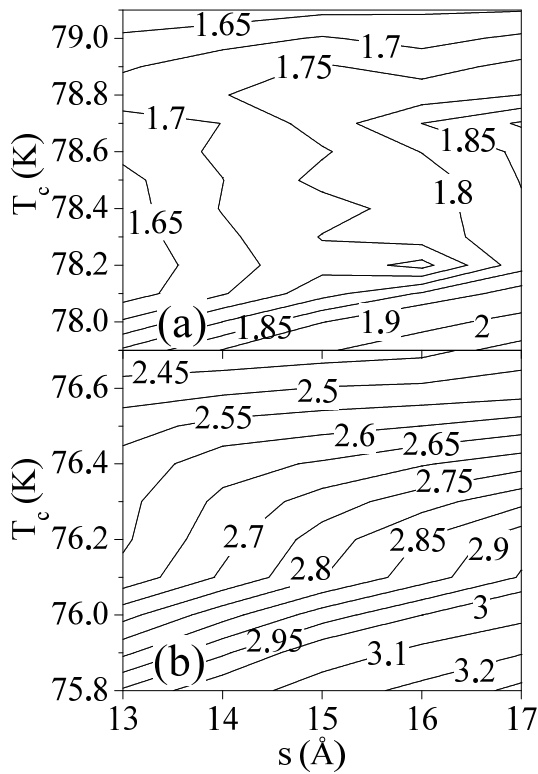}
\caption{ \label{fig_4} Contour plots of the quasiparticle scattering time $\tau$ in
the $(s, T_{c})$ plane for sample WI3B\_1 (a) and WI3C\_1 (b). Labels are in units of $10^{-14}$ s.}
\end{figure}

\begin{figure}[htbp]
\includegraphics[width=7 cm]{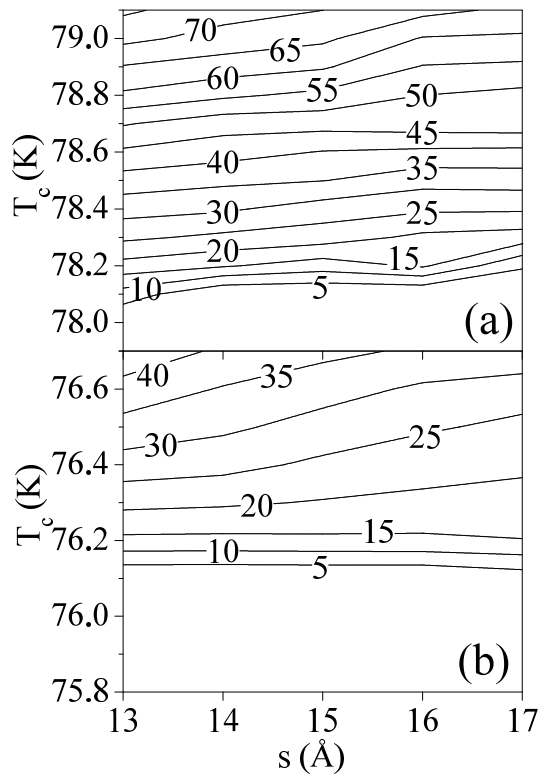}
\caption{ \label{fig_5} Contour plots of the hopping energy between adjacent layers $J$ in
the $(s, T_{c})$ plane for sample WI3B\_1 (a) and WI3C\_1 (b). Labels are in units of Kelvin degrees.}
\end{figure}

In Figs.~\ref{fig_3}, \ref{fig_4} and \ref{fig_5} the contour plots are presented for the best fit values of
$\tau_{\phi}$, $\tau$ and $J$, respectively.
These results confirm the relatively low dependence on the value of $s$;
the dependence on $T_{c}$ is also rather weak for $\tau_{\phi}$ and $\tau$, while $J$ shows
much stronger variations.

\begin{figure}
\includegraphics[width=7  cm]{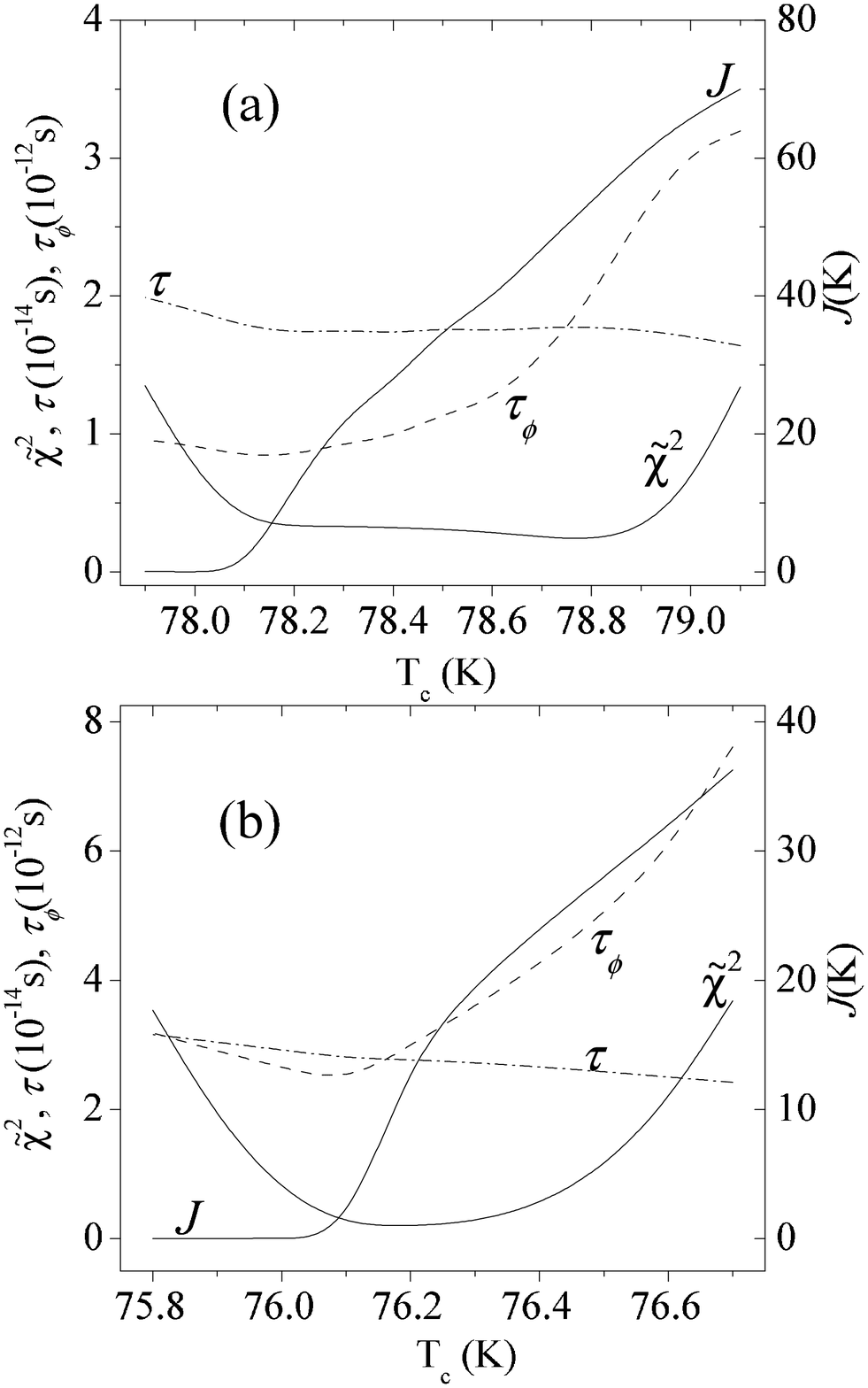}
\caption{ \label{fig_6} Behaviour of $\tilde \chi^{2}$, $\tau_{\phi}$, $\tau$ and $J$ along the $s=$15 \AA~section
of the $(s, T_{c})$ plane for sample WI3B\_1 (a) and WI3C\_1 (b). $\tilde \chi^{2}$, $\tau_{\phi}$,
$\tau$ values are referred to the left axis, $J$ is referred to the right axis.}
\end{figure}

It is worth stressing that the fit results are very similar
for both samples, confirming the reliability of the analysis.
Because of the minor influence of the $s$ parameter, we subsequently decided to fix $s=$15 \AA, as
expected from the crystal structure.
The corresponding behaviours for all the relevant quantities are compared in Fig.~\ref{fig_6}
as a function of $T_{c}$.

\begin{figure}[htbp]
\includegraphics[angle=270,width=8.4cm]{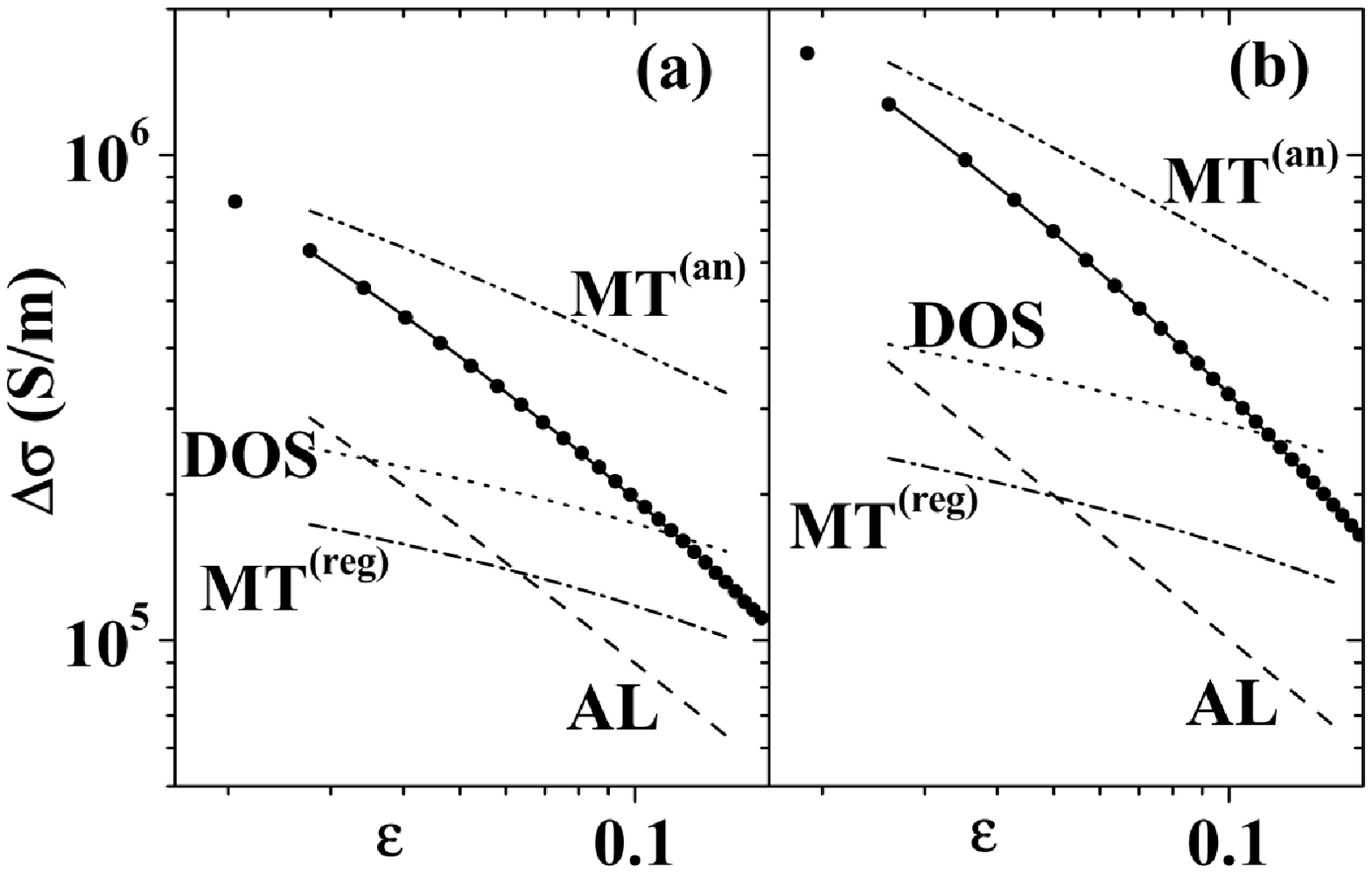}
\caption{ \label{fig_7} Best fits of experimental data to Eq.~(\ref{DS_total}) for sample WI3B\_1 (a) and WI3C\_1 (b).
Fits are represented by solid lines, single contributions by dashed and dotted lines, as indicated. Negative terms
(i.e. DOS and $\text{MT}^\text{(reg)}$) are plotted in absolute values.  The corresponding parameter values are
reported in Table~\ref{Tab_2}.}
\end{figure}

The $\tilde \chi^{2}$ minima can be clearly appreciated.
We defined a $\tilde \chi^{2}$ minimum as the region where $\tilde \chi^{2} \le 0.5$.
Therefore, the $T_{c}$ of each sample was defined as the center of the $\tilde \chi^{2}$ minimum and
the corresponding uncertainty was assumed as its half-width. This resulted in $T_{c}=78.5\pm0.4$ K and
$T_{c}=76.2 \pm0.1$ K for sample WI3B\_1 and WI3C\_1, respectively, which are both within the $T_{c}$ ranges
estimated in a preliminary way by means of the pure 2D AL law.
The estimates of $\tau_{\phi}$, $\tau$ and $J$ were defined as the parameter values corresponding to the $T_{c}$
evaluated for each sample, while the respective uncertainties were determined as
the maximum deviations from the estimated values throughout the whole uncertainty range allowed for $T_{c}$.
These results are summarized in Table~\ref{Tab_2} and the corresponding best fits are shown in Fig.~\ref{fig_7}, 
together with the measured $\Delta \sigma$ data reported as a function of $\epsilon$.

\begin{table}[htbp]  
\caption{\label{Tab_2} Best fit parameters deduced from fitting the experimental data of both samples to 
Eq.~(\ref{DS_total}).}
\begin{ruledtabular}
\begin{tabular}{ccccc}
Sample       &   $T_{c}$(K)     &  $\tau_\phi$  $\left( 10^{-12} \mbox{ s}\right)$ & $\tau$  $\left( 10^{-14} \mbox{ s} \right)$  &      $J$(K)         \\  \hline
WI3B\_1      &   78.5$\pm$0.4   &             1.1$^{+1.5}_{-0.3}$                  &          1.77$^{+0.01}_{-0.04}$              &  35$^{+25}_{-35}$   \\
WI3C\_1      &   76.2$\pm$0.1   &             3.0$\pm$0.6                          &          2.77$^{+0.03}_{-0.05}$              &  14$^{+6}_{-14}$    \\
\end{tabular}
\end{ruledtabular}
\end{table}

From the point of view of the experimental results, it should be noted that, as far as comparisons are possible, 
our data show the same qualitative behaviour already reported for Bi-2212 whiskers,\cite{Latyshev_93} single crystals 
\cite{Han_1997,Han_1998,Chowdhury_1999,Silva_2001} and thin films,\cite{Balestrino_1992} and for YBa$_{2}$Cu$_{2}$O$_{7 -\delta}$ (YBCO)
epitaxial thin films as well.\cite{Hopfengartner_1991} In particular, they share with previous results the order of magnitude 
of $\Delta \sigma$ near $T_{c}$ \cite{Pomar_1996,Chowdhury_1999,Silva_2001,Kim_2003}, the fact that $\Delta \sigma$ is sample dependent 
\cite{Freitas_1987,Hopfengartner_1991,Latyshev_93,Holm_1995,Pomar_1996,Han_1997,Han_1998,Chowdhury_1999,Silva_2001,Bjornangen_2001} and 
the same dependence on the cross section already shown in Bi-2212 whiskers.\cite{Latyshev_93}

From the point of view of the fits, 
it is apparent that the theory excellently fits the data of both samples in a rather large reduced temperature range:
$0.021 \leq \varepsilon\ \leq 0.148$ for sample WI3B\_1 and $0.026 \leq \varepsilon\ \leq 0.155$ for sample WI3C\_1.
About the parameter values, we note that $J$ and $\tau$ are consistent with previous reports for Bi-2212,
\cite{Thopart_2000,Nygmatulin_1996,Chowdhury_1999,Heine_1999,Balestrino_2001,Wahl_1999_2}
while the $\tau_\phi$ values are, to our knowledge, the largest ever observed on any HTSC compound.
\cite{Thopart_2000,Nygmatulin_1996,Chowdhury_1999,Heine_1999,Balestrino_2001,Wahl_1999_2,Bjornangen_2001,
Kim_2003,Holm_1995,Axnas_1998,Latyshev_1995,Wahl_1999,Semba_1997}
Correspondingly, the anomalous MT terms
exceed all the other contributions of at least a factor 2 over the whole temperature range, therefore
reassessing the commonly accepted hierarchy.

In order to clarify the reason for this discrepancy with the previous literature, we subsequently supposed the 
Maki-Thomson contribution to
be negligible and tried to fit the experimental data to the truncated formula:
\begin{equation}
\Delta \sigma = \Delta \sigma_{\text{AL}} + \Delta \sigma_{\text{DOS}} \quad , \label{DS_trunc}
\end{equation}
while retaining the same definitions for all the other quantities, included $r$ and $k$, which are valid 
for the general \textit{intermediate} case.

\begin{figure}[htbp]
\includegraphics[angle=270,width=8.4cm]{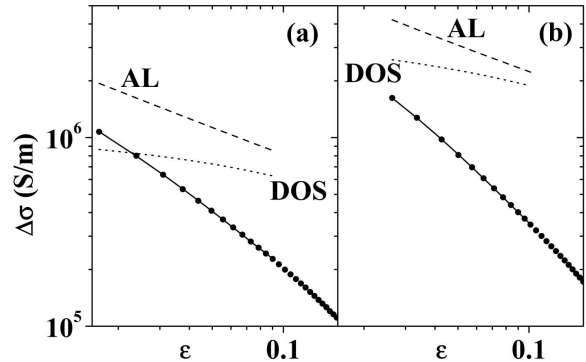}
\caption{ \label{fig_8} Best fits of experimental data to Eq.~(\ref{DS_trunc}) in the general \textit{intermediate}
case for sample WI3B\_1 (a) and WI3C\_1 (b).
Fits are represented by solid lines, single contributions by dashed and dotted lines, as indicated. The DOS 
term is plotted in absolute value. The corresponding parameter values are
reported in Table~\ref{Tab_3}.}
\end{figure}

The preliminary trials performed with the previously described fitting procedure showed in this case 
a much stronger dependence of $\tilde \chi ^{2}$ on the $s$ values and pointed out that no reasonable value of  
$\tilde \chi ^{2}$ was achievable neither in the $s \approx 15$ \AA~nor in the $s \approx 5$ \AA~region.
Therefore we slightly modified our procedure by fixing \textit{a priori} only $T_{c}$ and allowing $s$ 
to vary as a free parameter, which did not increase the number of free parameters with respect to the full 
theory case because of the disappearing of $\tau_\phi$. As a consequence, in this case only a scanning of the $T_{c}$
axis was performed by fixing its values in the range indicated by the pure 2D AL law. 
In this way we found that it was actually possible to obtain a good 
fit for the experimental results in the reduced temperature ranges $0.017 \leq \varepsilon\ \leq 0.09$ and 
$0.026 \leq \varepsilon\ \leq 0.10$ for sample WI3B\_1 and sample WI3C\_1, respectively.
These fits are shown in Fig.~\ref{fig_8} for the best possible choices of $T_{c}$ and the corresponding values of 
$\tilde \chi^{2}$ and of the other parameters are listed in Table~\ref{Tab_3}.

\begin{table}[htbp]  
\caption{\label{Tab_3} Best fit parameters corresponding to the fit of experimental data of both samples to 
Eq.~(\ref{DS_trunc}) in the general \textit{intermediate} case. 
$\tilde \chi ^{2}=\chi ^{2}/\text{d.o.f.}$, where $\text{d.o.f.}=10$ for sample WI3B\_1 and $\text{d.o.f.}=9$ 
for sample WI3C\_1.}
\begin{ruledtabular}
\begin{tabular}{cccccc}
Sample   & $T_{c}$(K)  &  $\tilde \chi^{2}$ &  $s$ $\left( \mbox{\AA}\right)$ & $\tau$  $\left( 10^{-14} \mbox{ s} \right)$ & $J$(K)         \\  \hline
WI3B\_1  &   78.24    &   0.183        &   0.45     &   0.56      &  616  \\
WI3C\_1  &   75.63    &   0.169        &   0.15     &   1.43      &  500  \\
\end{tabular}
\end{ruledtabular}
\end{table}

It is clear from Fig.~\ref{fig_8} that for this theoretical law the DOS contribution is significant, 
ranging from 45\% to 84\% of the AL term thoughout the fitted temperature range for the two samples.
Table~\ref{Tab_3} shows however that in this approximation 
reasonable values of $\tilde \chi^{2}$ and $\tau$ are accompanied by
very problematic values of $s$ and $J$. In fact, to our knowledge, only Bjorn$\ddot{\text{a}}$ngen \textit{et al.}
\cite{Bjornangen_2001} have reported compatible values for $J$ and no study has ever obtained similar values of $s$,
which appear to be 
very hard to reconcile with the Bi-2212 crystal structure. In our opinion, this implies that this 
description of the paraconductivity behaviour is not acceptable.

\begin{figure}[htbp]
\includegraphics[angle=270,width=8.4cm]{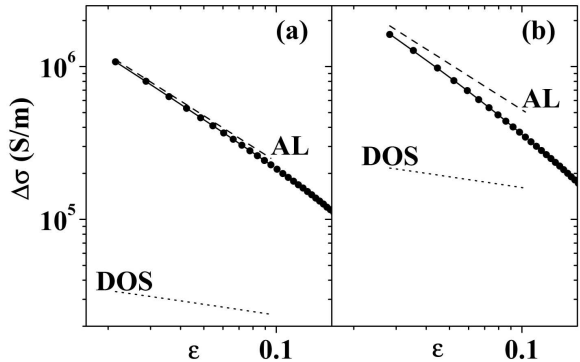}
\caption{ \label{fig_9} Best fits of experimental data to Eq.~(\ref{DS_trunc}) in the \textit{clean limit} case 
for sample WI3B\_1 (a) and WI3C\_1 (b).
Fits are represented by solid lines, single contributions by dashed and dotted lines, as indicated. The DOS 
term is plotted in absolute value. The corresponding parameter values are
reported in Table~\ref{Tab_4}.}
\end{figure}

As a further step in bridging the gap between previous results and ours, we tried to fit the experimental data to 
the truncated formula of Eq.~(\ref{DS_trunc}) in the \textit{clean limit}, which implies to change the general 
definitions of $r$ and $k$ for the intermediate case into: 
\begin{eqnarray}
r(T) & = & \frac{7 \zeta(3)J^{2}}{8\pi^{2}(k_BT)^2}  \label{r_clean}  \quad \quad \text{and} \\
k(T) & = & \frac{8\pi^{2}(k_BT\tau)^2 }{7 \zeta(3)\hbar^2  } \label{k_clean} \ , 
\end{eqnarray}

respectively, where $\zeta(x)$ is the Riemann zeta function. The absolute temperature $T$ range and the procedure of the fit 
were exactly the same as for Eq.~(\ref{DS_trunc}) in the intermediate case, 
which has been shown in Fig.~\ref{fig_8}.   
Fits with reasonable values of $\tilde \chi^{2}$ could be achieved in this limit, too. 
The parameter values and the curves corresponding to the best possible choice of $T_{c}$  
are displayed in Table~\ref{Tab_4} and Fig.~\ref{fig_9}, respectively.

\begin{table}[htbp]  
\caption{\label{Tab_4} Best fit parameters corresponding to the fit of experimental data of both samples to 
Eq.~(\ref{DS_trunc}) in the \textit{clean limit} case. 
$\tilde \chi ^{2}=\chi ^{2}/\text{d.o.f.}$, where $\text{d.o.f.}=10$ for sample WI3B\_1 and $\text{d.o.f.}=9$ 
for sample WI3C\_1.}
\begin{ruledtabular}
\begin{tabular}{cccccc}
Sample   & $T_{c}$(K)  &  $\tilde \chi^{2}$ &  $s$ $\left( \mbox{\AA}\right)$ & $\tau$  $\left( 10^{-14} \mbox{ s} \right)$ & $J$(K)         \\  \hline
WI3B\_1  &   77.86    &   0.142        &   6.4     &   0.95      &  1.6  \\
WI3C\_1  &   75.49    &   0.561        &   2.9     &   1.73      &  1.2  \\
\end{tabular}
\end{ruledtabular}
\end{table}

It is apparent that all the values in Table~\ref{Tab_4} seem to be acceptable in this case. In particular,
the $J$ values are compatible at least with others reported for Bi-2212 
single crystals,\cite{Heine_1999} for (Tl,Hg)$_{2}$Ba$_{2}$Ca$_{2}$Cu$_{3}$O$_{10+\delta}$  
single crystals \cite{Wahl_1999} and for Tl$_{2}$Ba$_{2}$CaCu$_{2}$O$_{8+y}$ single crystals. \cite{Kim_2003}
Moreover, it should be considered that the scanning of the $T_{c}$ axis revealed that in this case
$J$ was the parameter most sensitive to the choice of $T_{c}$, describing $J$ vs $T_{c}$ curves very similar to
the ones shown in Fig.~\ref{fig_6}, so that the values reported in Table~\ref{Tab_4} should be considered as  
lower limits for the $J$ values, the upper ones being about 20 K and 27 K for sample WI3B\_1 and WI3C\_1, 
respectively. This makes such results agree also with some others already reported 
for Bi-2212. \cite{Wahl_1999_2,Thopart_2000,Balestrino_2001}
As far $s$ is concerned, the results appear to be sample dependent and resembling the ones that can be extracted by
applying the pure 2D AL formula near $T_{c}$, as it was done in the inset of Fig.~\ref{fig.1_PRB}. 
In spite of the fact that they have no definite correspondence with the crystal structure, they 
agree with the values reported in previous studies of Bi-2212 thin films \cite{Balestrino_2001} and whiskers. 
\cite{Latyshev_93} 
The values for $\tau$ listed in Table~\ref{Tab_4} are very common and reasonable, too, and the combination 
of these parameters results in a remarkable decrease of the importance of the DOS term with respect 
to the intermediate case: for the clean limit approximation the DOS contribution ranges only 
between 10\% and 32\% of the AL term for the two samples.   
Therefore it could be stated that the AL process is dominant and that the DOS term is only 
a correction, with no need of the Maki-Thompson process to explain the experimental data, which is the 
hierarchy commonly accepted between the various paraconductivity contributions.

Nevertheless, it is very important to stress that the only difference between Fig.~\ref{fig_8} and Fig.~\ref{fig_9}
lies in the definition of the coefficients $r$ and $k$. In order to be allowed to use the definitions of
Eqs.~(\ref{r_clean})-(\ref{k_clean}) one must be in the clean limit, which means 
$4 \pi \tau k_{B} T / \hbar \gg 1$. Using the results listed in Table~\ref{Tab_4}, it can be calculated that   
$4 \pi \tau k_{B} T / \hbar =$ 1.24--1.34 for sample WI3B\_1 and $4 \pi \tau k_{B} T / \hbar =$ 2.21--2.38 
for sample WI3C\_1, over the temperature range used for the analysis.
These results are quite far from the clean limit and imply that it cannot be applied in a self-consistent
way to the interpretation of the data. Therefore this intepretation has to be rejected, so that the 
only satisfactory way to account for the data is the full theory in the intermediate case represented 
by Eqs.~(\ref{DS_total})-(\ref{MT_reg}), Fig.~\ref{fig_7} and Table~\ref{Tab_2}.

At the same time, the comparison between Fig.~\ref{fig_8} and Fig.~\ref{fig_9} clarifies the important role played 
by the coefficient $k$ in the theory in order to discern the importance of the different processes.
Consequently, a similar influence for the coefficient $\tilde k$ can be suspected and for this reason 
we repeated the fits for the full theory in the intermediate case using the $T_{c}$ values corresponding to the
$\tilde \chi^{2}$ minima listed in Table~\ref{Tab_2}. The procedure was exactly the same used to obtain also the 
other parameter values reported in Table~\ref{Tab_2}, 
except for the definition of $\tilde k$, which followed the formula published in Ref.~\onlinecite{Varlamov_1999}.
A $\tau_{\phi}$ decrease by 38\% was observed in both samples, together with an increase of $\tau$ (29--38\%) 
and a decrease of $J$ (20--50\%),
but these changes are not large enough to significantly modify the hierarchy between the contributions: the
anomalous MT terms are only slightly reduced (less than 5\%), the AL terms increased by less than 4\% and the
DOS term increase, even if important (32--35\%), does not allow to equal the anomalous MT contribution.
The most important effect in using the formula of Ref.~\onlinecite{Varlamov_1999} was the reduction of 
the regular MT terms, which were
depressed by 62--69\%, as expected both from the additional 1/2 factor and from the increase 
of $\tau$ in the $\psi ''(1/2)$ coefficient of $\tilde k$. The fits showed that almost all of 
this reduction was counterbalanced by the only other negative term, i.e. the DOS term, which increased accordingly.
Therefore the correction in the definition of the coefficient $\tilde k$ seems to play only a
minor role in our results. 

Once ascertained the correct interpretation of the data, 
in order to crosscheck our results, we computed $\xi_{c}(0)=s\sqrt{ r\left( T_{c} \right)}/2$, obtaining
$\xi_{c}(0)=0.8$ \AA~and $\xi_{c}(0)=0.4$ \AA~for samples WI3B\_1 and WI3C\_1, respectively.
Such values are in general agreement with previous experiments 
on Bi-2212.\cite{Han_1997,Palstra_1988,Pomar_1996,Mun_1993}
Also the Fermi level $E_{F}$ can be estimated in the
Drude picture from $\rho^{n}_{a}$ and the
microscopic parameters, via the formula $E_{F}=\hbar^{2}2\pi s/(e^{2}\tau\rho^{n}_{a})$.
This gives 1.4 eV and 1.1 eV for samples WI3B\_1 and WI3C\_1, respectively, which have to be considered
as upper limits of the real values, because of the assumption of a cylindrical Fermi surface underlying the theory.
Nevertheless, these energies agree fairly well with a previous analysis of the same kind
on thin films.\cite{Nygmatulin_1996}
Moreover, the same picture provides carrier densities $n=m^{\ast}/(e^{2}\tau\rho^{n}_{a})$ about
1.5--1.9 $\cdot 10^{21}$ cm$^{-3}$ for the two samples, if the bare electron
mass is assumed for the effective carrier mass $m^{\ast}$. This estimate lies in the same
range of a determination by chemical methods \cite{Idemoto_1995}
and within a factor of 2--3 from the results obtained by
Hall coefficient measurements.\cite{Hopfengartner_1993,Iye_1989,Maeda_1989}.

The results of our analysis
listed in Table~\ref{Tab_2}
imply that $4 \pi \tau k_{B} T / \hbar$=2.3--4.0, confirming that the clean limit
$4 \pi \tau k_{B} T / \hbar \gg 1$ is not fully developed for these systems.
We also note that $\tau_\phi / \tau \approx$60--110, therefore fulfilling the requirement
$\tau_\phi > \tau$ for the inelastic vs elastic scattering and showing that about one hundred scattering events
are required before the MT pairs are destroyed.
The comparison between the phase-breaking $\gamma_{\phi}$ and the anisotropy $r$ parameters
shows that $\gamma_{\phi}/r \approx $ 1.2--1.8 and therefore neither the weak ($\ll 1$) nor the strong ($\gg 1$)
limit can be considered for the pair breaking intensity.
We obtain for the interlayer tunnelling rates
$J^{2}\tau / \hbar^{2}\approx$ 0.9--3.7 $\cdot 10^{11} \ \text{s}^{-1}$ and for
the phase-breaking rates $\tau_{\phi}^{-1} \approx$ 3.3--8.7 $\cdot 10^{11} \ \text{s}^{-1}$.
Therefore, the ratio between the two rates, $\tau_{\phi}^{-1} /(J^{2}\tau / \hbar^{2})$, is about 2.3--3.6, so that
the condition for incoherent tunnelling along the
\textit{c}-axis ($\tau_{\phi}^{-1} /(J^{2}\tau / \hbar^{2})>1$) holds, although only marginally.
Consequently, Josephson effect precursor phenomena
cannot be excluded \textit{a priori}.
Finally, we note that these results set the \textit{s}-wave symmetry as an important component of
the order parameter for $T \gtrsim T_{c}$.
This is the same conclusion already achieved for $T \leq  T_{c}$ by recent experiments
on the \textit{c}-axis Josephson tunnelling across twisted Bi-2212 bicrystals.\cite{Li_1999}
Furthermore, our observation of a marginally incoherent tunnelling in the \textit{c}-axis direction could
support the Klemm's interpretation\cite{Klemm_2003} of another twist tunnelling experiment\cite{Takano_2002} in
overdoped Bi-2212 whiskers, according to which the results by Takano \textit{et al.}\cite{Takano_2002}
are consistent with an \textit{s}-wave symmetry accompanied by an intrinsically
coherent \textit{c}-axis tunnelling and a hot-spot Fermi surface.

\section{\label{Con}CONCLUSIONS}
The high quality of the experimental data and the careful fitting procedure allowed us to show
that the MT process can be the most important contribution to the superconducting
fluctuations above $T_{c}$ in real systems and that microscopic parameters can be reliably extracted
from the study of the paraconductivity.
The inconsistency of more traditional alternative interpretations was also demonstrated. 
This gives an indication for an \textit{s}-wave symmetry of the order parameter in Bi-2212 at $T \gtrsim  T_{c}$.
The revaluation of the MT process could also provide a physical mechanism for the appearance in 
underdoped samples of a
single-particle excitation energy $\Delta_{p}$ considerably greater than the coherence energy range $\Delta_{c}$ of
the condensed phase.\cite{Deutscher_1999}


%



\begin{acknowledgments}

This work was partly supported by INFM under project PAIS-STRIPES. We would like to acknowledge
C. Manfredotti, E. Vittone, A. Lo Giudice, E. Silva, R. Fastampa, S. Sarti and R. Gonnelli for
useful discussions. Special thanks to A. Varlamov for his valuable help.

\end{acknowledgments}

\bibliography{paracond}

\begin{thebibliography}{46}
\expandafter\ifx\csname natexlab\endcsname\relax\def\natexlab#1{#1}\fi
\expandafter\ifx\csname bibnamefont\endcsname\relax
  \def\bibnamefont#1{#1}\fi
\expandafter\ifx\csname bibfnamefont\endcsname\relax
  \def\bibfnamefont#1{#1}\fi
\expandafter\ifx\csname citenamefont\endcsname\relax
  \def\citenamefont#1{#1}\fi
\expandafter\ifx\csname url\endcsname\relax
  \def\url#1{\texttt{#1}}\fi
\expandafter\ifx\csname urlprefix\endcsname\relax\def\urlprefix{URL }\fi
\providecommand{\bibinfo}[2]{#2}
\providecommand{\eprint}[2][]{\url{#2}}

\bibitem[{\citenamefont{Aslamazov and Larkin}(1968)}]{AL_1968}
\bibinfo{author}{\bibfnamefont{L.~G.} \bibnamefont{Aslamazov}}
  \bibnamefont{and} \bibinfo{author}{\bibfnamefont{A.~I.}
  \bibnamefont{Larkin}}, \bibinfo{journal}{Phys. Lett.}
  \textbf{\bibinfo{volume}{26A}}, \bibinfo{pages}{238} (\bibinfo{year}{1968}).

\bibitem[{\citenamefont{Maki}(1968)}]{Maki_1968}
\bibinfo{author}{\bibfnamefont{K.}~\bibnamefont{Maki}},
  \bibinfo{journal}{Progr. Theor. Phys.} \textbf{\bibinfo{volume}{40}},
  \bibinfo{pages}{193} (\bibinfo{year}{1968}).

\bibitem[{\citenamefont{Thompson}(1970)}]{Thompson_1970}
\bibinfo{author}{\bibfnamefont{R.~S.} \bibnamefont{Thompson}},
  \bibinfo{journal}{Phys. Rev. B} \textbf{\bibinfo{volume}{1}},
  \bibinfo{pages}{327} (\bibinfo{year}{1970}).

\bibitem[{\citenamefont{Skocpol and Tinkham}(1975)}]{Skocpol_Tinkham_1975}
\bibinfo{author}{\bibfnamefont{W.~J.} \bibnamefont{Skocpol}} \bibnamefont{and}
  \bibinfo{author}{\bibfnamefont{M.}~\bibnamefont{Tinkham}},
  \bibinfo{journal}{Rep. Prog. Phys.} \textbf{\bibinfo{volume}{38}},
  \bibinfo{pages}{1049} (\bibinfo{year}{1975}).

\bibitem[{\citenamefont{Dorin et~al.}(1993)\citenamefont{Dorin, Klemm,
  Varlamov, Budzin, and Livanov}}]{Dorin_1993}
\bibinfo{author}{\bibfnamefont{V.~V.} \bibnamefont{Dorin}},
  \bibinfo{author}{\bibfnamefont{R.~A.} \bibnamefont{Klemm}},
  \bibinfo{author}{\bibfnamefont{A.~A.} \bibnamefont{Varlamov}},
  \bibinfo{author}{\bibfnamefont{A.~I.} \bibnamefont{Budzin}},
  \bibnamefont{and} \bibinfo{author}{\bibfnamefont{D.~V.}
  \bibnamefont{Livanov}}, \bibinfo{journal}{Phys. Rev. B}
  \textbf{\bibinfo{volume}{48}}, \bibinfo{pages}{12951} (\bibinfo{year}{1993}).

\bibitem[{\citenamefont{Varlamov et~al.}(1999)\citenamefont{Varlamov,
  Balestrino, Milani, and Livanov}}]{Varlamov_1999}
\bibinfo{author}{\bibfnamefont{A.~A.} \bibnamefont{Varlamov}},
  \bibinfo{author}{\bibfnamefont{G.}~\bibnamefont{Balestrino}},
  \bibinfo{author}{\bibfnamefont{E.}~\bibnamefont{Milani}}, \bibnamefont{and}
  \bibinfo{author}{\bibfnamefont{D.~V.} \bibnamefont{Livanov}},
  \bibinfo{journal}{Adv. Phys.} \textbf{\bibinfo{volume}{48}},
  \bibinfo{pages}{655} (\bibinfo{year}{1999}).

\bibitem[{\citenamefont{Larkin and Varlamov}(to be published)}]{Larkin_2004}
\bibinfo{author}{\bibfnamefont{A.~I.} \bibnamefont{Larkin}} \bibnamefont{and}
  \bibinfo{author}{\bibfnamefont{A.~A.} \bibnamefont{Varlamov}},
  \emph{\bibinfo{title}{Theory of Fluctuations in Superconductors}}
  (\bibinfo{publisher}{Oxford University Press}, \bibinfo{address}{Oxford},
  \bibinfo{year}{to be published}).

\bibitem[{\citenamefont{Axn$\ddot{\text{a}}$s
  et~al.}(1998)\citenamefont{Axn$\ddot{\text{a}}$s, Lundqvist, and
  $\ddot{\text{O}}$. Rapp}}]{Axnas_1998}
\bibinfo{author}{\bibfnamefont{J.}~\bibnamefont{Axn$\ddot{\text{a}}$s}},
  \bibinfo{author}{\bibfnamefont{B.}~\bibnamefont{Lundqvist}},
  \bibnamefont{and} \bibinfo{author}{\bibnamefont{$\ddot{\text{O}}$. Rapp}},
  \bibinfo{journal}{Phys. Rev. B} \textbf{\bibinfo{volume}{58}},
  \bibinfo{pages}{6628} (\bibinfo{year}{1998}).

\bibitem[{\citenamefont{Wahl et~al.}(1999{\natexlab{a}})\citenamefont{Wahl,
  Thopart, Villard, Maignan, Hardy, Soret, Ammor, and Ruyter}}]{Wahl_1999}
\bibinfo{author}{\bibfnamefont{A.}~\bibnamefont{Wahl}},
  \bibinfo{author}{\bibfnamefont{D.}~\bibnamefont{Thopart}},
  \bibinfo{author}{\bibfnamefont{G.}~\bibnamefont{Villard}},
  \bibinfo{author}{\bibfnamefont{A.}~\bibnamefont{Maignan}},
  \bibinfo{author}{\bibfnamefont{V.}~\bibnamefont{Hardy}},
  \bibinfo{author}{\bibfnamefont{J.~C.} \bibnamefont{Soret}},
  \bibinfo{author}{\bibfnamefont{L.}~\bibnamefont{Ammor}}, \bibnamefont{and}
  \bibinfo{author}{\bibfnamefont{A.}~\bibnamefont{Ruyter}},
  \bibinfo{journal}{Phys. Rev. B} \textbf{\bibinfo{volume}{59}},
  \bibinfo{pages}{7216} (\bibinfo{year}{1999}{\natexlab{a}}).

\bibitem[{\citenamefont{Wahl et~al.}(1999{\natexlab{b}})\citenamefont{Wahl,
  Thopart, Villard, Maignan, Simon, Soret, Ammor, and Ruyter}}]{Wahl_1999_2}
\bibinfo{author}{\bibfnamefont{A.}~\bibnamefont{Wahl}},
  \bibinfo{author}{\bibfnamefont{D.}~\bibnamefont{Thopart}},
  \bibinfo{author}{\bibfnamefont{G.}~\bibnamefont{Villard}},
  \bibinfo{author}{\bibfnamefont{A.}~\bibnamefont{Maignan}},
  \bibinfo{author}{\bibfnamefont{C.}~\bibnamefont{Simon}},
  \bibinfo{author}{\bibfnamefont{J.~C.} \bibnamefont{Soret}},
  \bibinfo{author}{\bibfnamefont{L.}~\bibnamefont{Ammor}}, \bibnamefont{and}
  \bibinfo{author}{\bibfnamefont{A.}~\bibnamefont{Ruyter}},
  \bibinfo{journal}{Phys. Rev. B} \textbf{\bibinfo{volume}{60}},
  \bibinfo{pages}{12495} (\bibinfo{year}{1999}{\natexlab{b}}).

\bibitem[{\citenamefont{Thopart et~al.}(2000)\citenamefont{Thopart, Wahl,
  Warmont, Simon, Soret, Ammor, Ruyter, Buzdin, Varlamov, and
  de~Brion}}]{Thopart_2000}
\bibinfo{author}{\bibfnamefont{D.}~\bibnamefont{Thopart}},
  \bibinfo{author}{\bibfnamefont{A.}~\bibnamefont{Wahl}},
  \bibinfo{author}{\bibfnamefont{F.}~\bibnamefont{Warmont}},
  \bibinfo{author}{\bibfnamefont{C.}~\bibnamefont{Simon}},
  \bibinfo{author}{\bibfnamefont{J.~C.} \bibnamefont{Soret}},
  \bibinfo{author}{\bibfnamefont{L.}~\bibnamefont{Ammor}},
  \bibinfo{author}{\bibfnamefont{A.}~\bibnamefont{Ruyter}},
  \bibinfo{author}{\bibfnamefont{A.~I.} \bibnamefont{Buzdin}},
  \bibinfo{author}{\bibfnamefont{A.~A.} \bibnamefont{Varlamov}},
  \bibnamefont{and} \bibinfo{author}{\bibfnamefont{S.}~\bibnamefont{de~Brion}},
  \bibinfo{journal}{Phys. Rev. B} \textbf{\bibinfo{volume}{62}},
  \bibinfo{pages}{9721} (\bibinfo{year}{2000}).

\bibitem[{\citenamefont{Kim et~al.}(2003)\citenamefont{Kim, Chowdhury, Kang,
  Zang, and Lee}}]{Kim_2003}
\bibinfo{author}{\bibfnamefont{H.-J.} \bibnamefont{Kim}},
  \bibinfo{author}{\bibfnamefont{P.}~\bibnamefont{Chowdhury}},
  \bibinfo{author}{\bibfnamefont{W.~N.} \bibnamefont{Kang}},
  \bibinfo{author}{\bibfnamefont{D.-J.} \bibnamefont{Zang}}, \bibnamefont{and}
  \bibinfo{author}{\bibfnamefont{S.-I.} \bibnamefont{Lee}},
  \bibinfo{journal}{Phys. Rev. B} \textbf{\bibinfo{volume}{67}},
  \bibinfo{pages}{144502} (\bibinfo{year}{2003}).

\bibitem[{\citenamefont{Bjorn$\ddot{\text{a}}$ngen
  et~al.}(2001)\citenamefont{Bjorn$\ddot{\text{a}}$ngen, Axn$\ddot{\text{a}}$s,
  Eltsev, Rydh, and $\ddot{\text{O}}$. Rapp}}]{Bjornangen_2001}
\bibinfo{author}{\bibfnamefont{T.}~\bibnamefont{Bjorn$\ddot{\text{a}}$ngen}},
  \bibinfo{author}{\bibfnamefont{J.}~\bibnamefont{Axn$\ddot{\text{a}}$s}},
  \bibinfo{author}{\bibfnamefont{Y.}~\bibnamefont{Eltsev}},
  \bibinfo{author}{\bibfnamefont{A.}~\bibnamefont{Rydh}}, \bibnamefont{and}
  \bibinfo{author}{\bibnamefont{$\ddot{\text{O}}$. Rapp}},
  \bibinfo{journal}{Phys. Rev. B} \textbf{\bibinfo{volume}{63}},
  \bibinfo{pages}{224518} (\bibinfo{year}{2001}).

\bibitem[{\citenamefont{Chowdhury and Bhatia}(1999)}]{Chowdhury_1999}
\bibinfo{author}{\bibfnamefont{P.}~\bibnamefont{Chowdhury}} \bibnamefont{and}
  \bibinfo{author}{\bibfnamefont{S.~N.} \bibnamefont{Bhatia}},
  \bibinfo{journal}{Physica C} \textbf{\bibinfo{volume}{319}},
  \bibinfo{pages}{150} (\bibinfo{year}{1999}).

\bibitem[{\citenamefont{Carretta et~al.}(1996)\citenamefont{Carretta, Livanov,
  Rigamonti, and Varlamov}}]{Carretta_1996}
\bibinfo{author}{\bibfnamefont{P.}~\bibnamefont{Carretta}},
  \bibinfo{author}{\bibfnamefont{D.~V.} \bibnamefont{Livanov}},
  \bibinfo{author}{\bibfnamefont{A.}~\bibnamefont{Rigamonti}},
  \bibnamefont{and} \bibinfo{author}{\bibfnamefont{A.~A.}
  \bibnamefont{Varlamov}}, \bibinfo{journal}{Phys. Rev. B}
  \textbf{\bibinfo{volume}{54}}, \bibinfo{pages}{R9682} (\bibinfo{year}{1996}).

\bibitem[{\citenamefont{Gorlova and Timofeev}(1995)}]{Gorlova_95}
\bibinfo{author}{\bibfnamefont{I.~G.} \bibnamefont{Gorlova}} \bibnamefont{and}
  \bibinfo{author}{\bibfnamefont{V.~N.} \bibnamefont{Timofeev}},
  \bibinfo{journal}{Physica C} \textbf{\bibinfo{volume}{255}},
  \bibinfo{pages}{131} (\bibinfo{year}{1995}).

\bibitem[{\citenamefont{Timofeev and Gorlova}(1998)}]{Timofeev_98}
\bibinfo{author}{\bibfnamefont{V.~N.} \bibnamefont{Timofeev}} \bibnamefont{and}
  \bibinfo{author}{\bibfnamefont{I.~G.} \bibnamefont{Gorlova}},
  \bibinfo{journal}{Physica C} \textbf{\bibinfo{volume}{309}},
  \bibinfo{pages}{113} (\bibinfo{year}{1998}).

\bibitem[{\citenamefont{Li et~al.}(1994)\citenamefont{Li, Kes, Fu, Menovsky,
  and Franse}}]{Li_1994}
\bibinfo{author}{\bibfnamefont{T.~W.} \bibnamefont{Li}},
  \bibinfo{author}{\bibfnamefont{P.~H.} \bibnamefont{Kes}},
  \bibinfo{author}{\bibfnamefont{W.~T.} \bibnamefont{Fu}},
  \bibinfo{author}{\bibfnamefont{A.~A.} \bibnamefont{Menovsky}},
  \bibnamefont{and} \bibinfo{author}{\bibfnamefont{J.~J.~M.}
  \bibnamefont{Franse}}, \bibinfo{journal}{Physica C}
  \textbf{\bibinfo{volume}{224}}, \bibinfo{pages}{110} (\bibinfo{year}{1994}).

\bibitem[{\citenamefont{Truccato et~al.}(2002)\citenamefont{Truccato, Rinaudo,
  Manfredotti, Agostino, Benzi, Volpe, Paolini, and P.Olivero}}]{Truccato_02}
\bibinfo{author}{\bibfnamefont{M.}~\bibnamefont{Truccato}},
  \bibinfo{author}{\bibfnamefont{G.}~\bibnamefont{Rinaudo}},
  \bibinfo{author}{\bibfnamefont{C.}~\bibnamefont{Manfredotti}},
  \bibinfo{author}{\bibfnamefont{A.}~\bibnamefont{Agostino}},
  \bibinfo{author}{\bibfnamefont{P.}~\bibnamefont{Benzi}},
  \bibinfo{author}{\bibfnamefont{P.}~\bibnamefont{Volpe}},
  \bibinfo{author}{\bibfnamefont{C.}~\bibnamefont{Paolini}}, \bibnamefont{and}
  \bibinfo{author}{\bibnamefont{P.Olivero}}, \bibinfo{journal}{Supercond. Sci.
  Technol.} \textbf{\bibinfo{volume}{15}}, \bibinfo{pages}{1304}
  (\bibinfo{year}{2002}).

\bibitem[{\citenamefont{Esposito et~al.}(2000)\citenamefont{Esposito, Muzzi,
  Sarti, Fastampa, and Silva}}]{Esposito_2000}
\bibinfo{author}{\bibfnamefont{M.}~\bibnamefont{Esposito}},
  \bibinfo{author}{\bibfnamefont{L.}~\bibnamefont{Muzzi}},
  \bibinfo{author}{\bibfnamefont{S.}~\bibnamefont{Sarti}},
  \bibinfo{author}{\bibfnamefont{R.}~\bibnamefont{Fastampa}}, \bibnamefont{and}
  \bibinfo{author}{\bibfnamefont{E.}~\bibnamefont{Silva}}, \bibinfo{journal}{J.
  Appl. Phys.} \textbf{\bibinfo{volume}{88}}, \bibinfo{pages}{2724}
  (\bibinfo{year}{2000}).

\bibitem[{\citenamefont{Stupp et~al.}(1992)\citenamefont{Stupp, Lee,
  Giapintzakis, and Ginsberg}}]{Stupp_1992}
\bibinfo{author}{\bibfnamefont{S.~E.} \bibnamefont{Stupp}},
  \bibinfo{author}{\bibfnamefont{W.~C.} \bibnamefont{Lee}},
  \bibinfo{author}{\bibfnamefont{J.}~\bibnamefont{Giapintzakis}},
  \bibnamefont{and} \bibinfo{author}{\bibfnamefont{D.~M.}
  \bibnamefont{Ginsberg}}, \bibinfo{journal}{Phys. Rev. B}
  \textbf{\bibinfo{volume}{45}}, \bibinfo{pages}{3093} (\bibinfo{year}{1992}).

\bibitem[{\citenamefont{Holm et~al.}(1995)\citenamefont{Holm,
  $\ddot{\text{O}}$. Rapp, Johnson, and Helmersson}}]{Holm_1995}
\bibinfo{author}{\bibfnamefont{W.}~\bibnamefont{Holm}},
  \bibinfo{author}{\bibnamefont{$\ddot{\text{O}}$. Rapp}},
  \bibinfo{author}{\bibfnamefont{C.~N.~L.} \bibnamefont{Johnson}},
  \bibnamefont{and}
  \bibinfo{author}{\bibfnamefont{U.}~\bibnamefont{Helmersson}},
  \bibinfo{journal}{Phys. Rev. B} \textbf{\bibinfo{volume}{52}},
  \bibinfo{pages}{3748} (\bibinfo{year}{1995}).

\bibitem[{\citenamefont{Latyshev et~al.}(1995)\citenamefont{Latyshev, Laborde,
  and Monceau}}]{Latyshev_1995}
\bibinfo{author}{\bibfnamefont{Y.~I.} \bibnamefont{Latyshev}},
  \bibinfo{author}{\bibfnamefont{O.}~\bibnamefont{Laborde}}, \bibnamefont{and}
  \bibinfo{author}{\bibfnamefont{P.}~\bibnamefont{Monceau}},
  \bibinfo{journal}{Europhys. Lett.} \textbf{\bibinfo{volume}{29}},
  \bibinfo{pages}{495} (\bibinfo{year}{1995}).

\bibitem[{\citenamefont{Heine et~al.}(1999)\citenamefont{Heine, Lang, Wang, and
  Dou}}]{Heine_1999}
\bibinfo{author}{\bibfnamefont{G.}~\bibnamefont{Heine}},
  \bibinfo{author}{\bibfnamefont{W.}~\bibnamefont{Lang}},
  \bibinfo{author}{\bibfnamefont{X.~L.} \bibnamefont{Wang}}, \bibnamefont{and}
  \bibinfo{author}{\bibfnamefont{S.~X.} \bibnamefont{Dou}},
  \bibinfo{journal}{Phys. Rev. B} \textbf{\bibinfo{volume}{59}},
  \bibinfo{pages}{11179} (\bibinfo{year}{1999}).

\bibitem[{\citenamefont{Semba and Matsuda}(1997)}]{Semba_1997}
\bibinfo{author}{\bibfnamefont{K.}~\bibnamefont{Semba}} \bibnamefont{and}
  \bibinfo{author}{\bibfnamefont{A.}~\bibnamefont{Matsuda}},
  \bibinfo{journal}{Phys. Rev. B} \textbf{\bibinfo{volume}{55}},
  \bibinfo{pages}{11103} (\bibinfo{year}{1997}).

\bibitem[{\citenamefont{Balestrino et~al.}(2001)\citenamefont{Balestrino,
  Crisan, Livanov, Manokhin, and Milani}}]{Balestrino_2001}
\bibinfo{author}{\bibfnamefont{G.}~\bibnamefont{Balestrino}},
  \bibinfo{author}{\bibfnamefont{A.}~\bibnamefont{Crisan}},
  \bibinfo{author}{\bibfnamefont{D.~V.} \bibnamefont{Livanov}},
  \bibinfo{author}{\bibfnamefont{S.~I.} \bibnamefont{Manokhin}},
  \bibnamefont{and} \bibinfo{author}{\bibfnamefont{E.}~\bibnamefont{Milani}},
  \bibinfo{journal}{Physica C} \textbf{\bibinfo{volume}{355}},
  \bibinfo{pages}{135} (\bibinfo{year}{2001}).

\bibitem[{\citenamefont{Pomar et~al.}(1996)\citenamefont{Pomar, Ramallo,
  Mosqueira, Torr\'on, and Vidal}}]{Pomar_1996}
\bibinfo{author}{\bibfnamefont{A.}~\bibnamefont{Pomar}},
  \bibinfo{author}{\bibfnamefont{M.~V.} \bibnamefont{Ramallo}},
  \bibinfo{author}{\bibfnamefont{J.}~\bibnamefont{Mosqueira}},
  \bibinfo{author}{\bibfnamefont{C.}~\bibnamefont{Torr\'on}}, \bibnamefont{and}
  \bibinfo{author}{\bibfnamefont{F.}~\bibnamefont{Vidal}},
  \bibinfo{journal}{Phys. Rev. B} \textbf{\bibinfo{volume}{54}},
  \bibinfo{pages}{7470} (\bibinfo{year}{1996}).

\bibitem[{\citenamefont{Meingast et~al.}(1996)\citenamefont{Meingast, Junod,
  and Walker}}]{Meingast_1994}
\bibinfo{author}{\bibfnamefont{C.}~\bibnamefont{Meingast}},
  \bibinfo{author}{\bibfnamefont{A.}~\bibnamefont{Junod}}, \bibnamefont{and}
  \bibinfo{author}{\bibfnamefont{E.}~\bibnamefont{Walker}},
  \bibinfo{journal}{Physica C} \textbf{\bibinfo{volume}{272}},
  \bibinfo{pages}{106} (\bibinfo{year}{1996}).

\bibitem[{\citenamefont{Latyshev et~al.}(1993)\citenamefont{Latyshev, Gorlova,
  Nikitina, Antokhina, Zybtsev, Kukhta, and Timofeev}}]{Latyshev_93}
\bibinfo{author}{\bibfnamefont{Y.~I.} \bibnamefont{Latyshev}},
  \bibinfo{author}{\bibfnamefont{I.~G.} \bibnamefont{Gorlova}},
  \bibinfo{author}{\bibfnamefont{A.~M.} \bibnamefont{Nikitina}},
  \bibinfo{author}{\bibfnamefont{V.~U.} \bibnamefont{Antokhina}},
  \bibinfo{author}{\bibfnamefont{S.~G.} \bibnamefont{Zybtsev}},
  \bibinfo{author}{\bibfnamefont{N.~P.} \bibnamefont{Kukhta}},
  \bibnamefont{and} \bibinfo{author}{\bibfnamefont{V.~N.}
  \bibnamefont{Timofeev}}, \bibinfo{journal}{Physica C}
  \textbf{\bibinfo{volume}{216}}, \bibinfo{pages}{471} (\bibinfo{year}{1993}).

\bibitem[{\citenamefont{Han et~al.}(1997)\citenamefont{Han, Zhao, Gu, Russell,
  and Koshizuka}}]{Han_1997}
\bibinfo{author}{\bibfnamefont{S.~H.} \bibnamefont{Han}},
  \bibinfo{author}{\bibfnamefont{Y.}~\bibnamefont{Zhao}},
  \bibinfo{author}{\bibfnamefont{G.~D.} \bibnamefont{Gu}},
  \bibinfo{author}{\bibfnamefont{G.~J.} \bibnamefont{Russell}},
  \bibnamefont{and}
  \bibinfo{author}{\bibfnamefont{N.}~\bibnamefont{Koshizuka}},
  \bibinfo{journal}{Phys. Stat. Sol. B} \textbf{\bibinfo{volume}{203}},
  \bibinfo{pages}{189} (\bibinfo{year}{1997}).

\bibitem[{\citenamefont{Han et~al.}(1998)\citenamefont{Han, Eltsev, and
  $\ddot{\text{O}}$. Rapp}}]{Han_1998}
\bibinfo{author}{\bibfnamefont{S.~H.} \bibnamefont{Han}},
  \bibinfo{author}{\bibfnamefont{Y.}~\bibnamefont{Eltsev}}, \bibnamefont{and}
  \bibinfo{author}{\bibnamefont{$\ddot{\text{O}}$. Rapp}},
  \bibinfo{journal}{Phys. Rev. B} \textbf{\bibinfo{volume}{57}},
  \bibinfo{pages}{7510} (\bibinfo{year}{1998}).

\bibitem[{\citenamefont{E.Silva et~al.}(2001)\citenamefont{E.Silva, Sarti,
  Fastampa, and Giura}}]{Silva_2001}
\bibinfo{author}{\bibnamefont{E.Silva}},
  \bibinfo{author}{\bibfnamefont{S.}~\bibnamefont{Sarti}},
  \bibinfo{author}{\bibfnamefont{R.}~\bibnamefont{Fastampa}}, \bibnamefont{and}
  \bibinfo{author}{\bibfnamefont{M.}~\bibnamefont{Giura}},
  \bibinfo{journal}{Phys. Rev. B}  (\bibinfo{year}{2001}).

\bibitem[{\citenamefont{Balestrino et~al.}(1992)\citenamefont{Balestrino,
  Marinelli, Milani, Reggiani, Vaglio, and Varlamov}}]{Balestrino_1992}
\bibinfo{author}{\bibfnamefont{G.}~\bibnamefont{Balestrino}},
  \bibinfo{author}{\bibfnamefont{M.}~\bibnamefont{Marinelli}},
  \bibinfo{author}{\bibfnamefont{E.}~\bibnamefont{Milani}},
  \bibinfo{author}{\bibfnamefont{L.}~\bibnamefont{Reggiani}},
  \bibinfo{author}{\bibfnamefont{R.}~\bibnamefont{Vaglio}}, \bibnamefont{and}
  \bibinfo{author}{\bibfnamefont{A.~A.} \bibnamefont{Varlamov}},
  \bibinfo{journal}{Phys. Rev. B} \textbf{\bibinfo{volume}{46}},
  \bibinfo{pages}{14919} (\bibinfo{year}{1992}).

\bibitem[{\citenamefont{H$\ddot{\text{o}}$pfengartner
  et~al.}(1991)\citenamefont{H$\ddot{\text{o}}$pfengartner, Hensel, and
  Saemann-Ischenko}}]{Hopfengartner_1991}
\bibinfo{author}{\bibfnamefont{R.}~\bibnamefont{H$\ddot{\text{o}}$pfengartner}%
}, \bibinfo{author}{\bibfnamefont{B.}~\bibnamefont{Hensel}}, \bibnamefont{and}
  \bibinfo{author}{\bibfnamefont{G.}~\bibnamefont{Saemann-Ischenko}},
  \bibinfo{journal}{Phys. Rev. B} \textbf{\bibinfo{volume}{44}},
  \bibinfo{pages}{741} (\bibinfo{year}{1991}).

\bibitem[{\citenamefont{Freitas et~al.}(1987)\citenamefont{Freitas, Tsuei, and
  Plaskett}}]{Freitas_1987}
\bibinfo{author}{\bibfnamefont{P.~P.} \bibnamefont{Freitas}},
  \bibinfo{author}{\bibfnamefont{C.~C.} \bibnamefont{Tsuei}}, \bibnamefont{and}
  \bibinfo{author}{\bibfnamefont{T.~S.} \bibnamefont{Plaskett}},
  \bibinfo{journal}{Phys. Rev. B} \textbf{\bibinfo{volume}{36}},
  \bibinfo{pages}{833} (\bibinfo{year}{1987}).

\bibitem[{\citenamefont{Nygmatulin et~al.}(1996)\citenamefont{Nygmatulin,
  Varlamov, Livanov, Balestrino, and Milani}}]{Nygmatulin_1996}
\bibinfo{author}{\bibfnamefont{A.~S.} \bibnamefont{Nygmatulin}},
  \bibinfo{author}{\bibfnamefont{A.~A.} \bibnamefont{Varlamov}},
  \bibinfo{author}{\bibfnamefont{D.~V.} \bibnamefont{Livanov}},
  \bibinfo{author}{\bibfnamefont{G.}~\bibnamefont{Balestrino}},
  \bibnamefont{and} \bibinfo{author}{\bibfnamefont{E.}~\bibnamefont{Milani}},
  \bibinfo{journal}{Phys. Rev. B} \textbf{\bibinfo{volume}{53}},
  \bibinfo{pages}{3557} (\bibinfo{year}{1996}).

\bibitem[{\citenamefont{Palstra et~al.}(1988)\citenamefont{Palstra, Batlogg,
  van Dover, and Waszczak}}]{Palstra_1988}
\bibinfo{author}{\bibfnamefont{T.~T.~M.} \bibnamefont{Palstra}},
  \bibinfo{author}{\bibfnamefont{B.}~\bibnamefont{Batlogg}},
  \bibinfo{author}{\bibfnamefont{L.~F. S. R.~B.} \bibnamefont{van Dover}},
  \bibnamefont{and} \bibinfo{author}{\bibfnamefont{J.~V.}
  \bibnamefont{Waszczak}}, \bibinfo{journal}{Phys. Rev. B}
  \textbf{\bibinfo{volume}{38}}, \bibinfo{pages}{5102} (\bibinfo{year}{1988}).

\bibitem[{\citenamefont{Mun et~al.}(1993)\citenamefont{Mun, Lee, Salk, Shin,
  and Joo}}]{Mun_1993}
\bibinfo{author}{\bibfnamefont{M.-O.} \bibnamefont{Mun}},
  \bibinfo{author}{\bibfnamefont{S.-I.} \bibnamefont{Lee}},
  \bibinfo{author}{\bibfnamefont{S.-H.~S.} \bibnamefont{Salk}},
  \bibinfo{author}{\bibfnamefont{H.~J.} \bibnamefont{Shin}}, \bibnamefont{and}
  \bibinfo{author}{\bibfnamefont{M.~K.} \bibnamefont{Joo}},
  \bibinfo{journal}{Phys. Rev. B} \textbf{\bibinfo{volume}{48}},
  \bibinfo{pages}{6703} (\bibinfo{year}{1993}).

\bibitem[{\citenamefont{Idemoto et~al.}(1995)\citenamefont{Idemoto, Toda, and
  Fueki}}]{Idemoto_1995}
\bibinfo{author}{\bibfnamefont{Y.}~\bibnamefont{Idemoto}},
  \bibinfo{author}{\bibfnamefont{T.}~\bibnamefont{Toda}}, \bibnamefont{and}
  \bibinfo{author}{\bibfnamefont{K.}~\bibnamefont{Fueki}},
  \bibinfo{journal}{Physica C} \textbf{\bibinfo{volume}{249}},
  \bibinfo{pages}{123} (\bibinfo{year}{1995}).

\bibitem[{\citenamefont{Hopfeng$\ddot{\text{a}}$rtner
  et~al.}(1993)\citenamefont{Hopfeng$\ddot{\text{a}}$rtner, Leghissa,
  Kreiselmeyer, Holzapfel, Schmitt, and Saemann-Ischenko}}]{Hopfengartner_1993}
\bibinfo{author}{\bibfnamefont{R.}~\bibnamefont{Hopfeng$\ddot{\text{a}}$rtner}%
}, \bibinfo{author}{\bibfnamefont{M.}~\bibnamefont{Leghissa}},
  \bibinfo{author}{\bibfnamefont{G.}~\bibnamefont{Kreiselmeyer}},
  \bibinfo{author}{\bibfnamefont{B.}~\bibnamefont{Holzapfel}},
  \bibinfo{author}{\bibfnamefont{P.}~\bibnamefont{Schmitt}}, \bibnamefont{and}
  \bibinfo{author}{\bibfnamefont{G.}~\bibnamefont{Saemann-Ischenko}},
  \bibinfo{journal}{Phys. Rev. B} \textbf{\bibinfo{volume}{47}},
  \bibinfo{pages}{5992} (\bibinfo{year}{1993}).

\bibitem[{\citenamefont{Iye et~al.}(1989)\citenamefont{Iye, Nakamura, and
  Tamegai}}]{Iye_1989}
\bibinfo{author}{\bibfnamefont{Y.}~\bibnamefont{Iye}},
  \bibinfo{author}{\bibfnamefont{S.}~\bibnamefont{Nakamura}}, \bibnamefont{and}
  \bibinfo{author}{\bibfnamefont{T.}~\bibnamefont{Tamegai}},
  \bibinfo{journal}{Physica C} \textbf{\bibinfo{volume}{159}},
  \bibinfo{pages}{616} (\bibinfo{year}{1989}).

\bibitem[{\citenamefont{Maeda et~al.}(1989)\citenamefont{Maeda, Noda,
  Takebayashi, and Uchinokura}}]{Maeda_1989}
\bibinfo{author}{\bibfnamefont{A.}~\bibnamefont{Maeda}},
  \bibinfo{author}{\bibfnamefont{K.}~\bibnamefont{Noda}},
  \bibinfo{author}{\bibfnamefont{S.}~\bibnamefont{Takebayashi}},
  \bibnamefont{and}
  \bibinfo{author}{\bibfnamefont{K.}~\bibnamefont{Uchinokura}},
  \bibinfo{journal}{Physica C} \textbf{\bibinfo{volume}{162-164}},
  \bibinfo{pages}{1205} (\bibinfo{year}{1989}).

\bibitem[{\citenamefont{Li et~al.}(1999)\citenamefont{Li, Tsay, Suenaga, Klemm,
  Gu, and Koshizuka}}]{Li_1999}
\bibinfo{author}{\bibfnamefont{Q.}~\bibnamefont{Li}},
  \bibinfo{author}{\bibfnamefont{Y.~N.} \bibnamefont{Tsay}},
  \bibinfo{author}{\bibfnamefont{M.}~\bibnamefont{Suenaga}},
  \bibinfo{author}{\bibfnamefont{R.~A.} \bibnamefont{Klemm}},
  \bibinfo{author}{\bibfnamefont{G.~D.} \bibnamefont{Gu}}, \bibnamefont{and}
  \bibinfo{author}{\bibfnamefont{N.}~\bibnamefont{Koshizuka}},
  \bibinfo{journal}{Phys. Rev. Lett.} \textbf{\bibinfo{volume}{83}},
  \bibinfo{pages}{4160} (\bibinfo{year}{1999}).

\bibitem[{\citenamefont{Klemm}(2003)}]{Klemm_2003}
\bibinfo{author}{\bibfnamefont{R.~A.} \bibnamefont{Klemm}},
  \bibinfo{journal}{Phys. Rev. B} \textbf{\bibinfo{volume}{67}},
  \bibinfo{pages}{174509} (\bibinfo{year}{2003}).

\bibitem[{\citenamefont{Takano et~al.}(2002)\citenamefont{Takano, Hatano,
  Fukuyo, Ishii, Ohmori, Arisawa, Togano, and Tachiki}}]{Takano_2002}
\bibinfo{author}{\bibfnamefont{Y.}~\bibnamefont{Takano}},
  \bibinfo{author}{\bibfnamefont{T.}~\bibnamefont{Hatano}},
  \bibinfo{author}{\bibfnamefont{A.}~\bibnamefont{Fukuyo}},
  \bibinfo{author}{\bibfnamefont{A.}~\bibnamefont{Ishii}},
  \bibinfo{author}{\bibfnamefont{M.}~\bibnamefont{Ohmori}},
  \bibinfo{author}{\bibfnamefont{S.}~\bibnamefont{Arisawa}},
  \bibinfo{author}{\bibfnamefont{K.}~\bibnamefont{Togano}}, \bibnamefont{and}
  \bibinfo{author}{\bibfnamefont{M.}~\bibnamefont{Tachiki}},
  \bibinfo{journal}{Phys. Rev. B} \textbf{\bibinfo{volume}{65}},
  \bibinfo{pages}{140513} (\bibinfo{year}{2002}).

\bibitem[{\citenamefont{Deutscher}(1999)}]{Deutscher_1999}
\bibinfo{author}{\bibfnamefont{G.}~\bibnamefont{Deutscher}},
  \bibinfo{journal}{Nature} \textbf{\bibinfo{volume}{397}},
  \bibinfo{pages}{410} (\bibinfo{year}{1999}).

\end{thebibliography}

\end{document}